\documentclass[conference]{IEEEtran}
\usepackage[letterpaper, left=0.62in, right=0.62in, bottom=1in, top=0.75in]{geometry}
\IEEEoverridecommandlockouts
\usepackage[binary-units]{siunitx}
\usepackage[caption=false,font=footnotesize]{subfig}
\usepackage{tabularx}
\usepackage{multirow}
\usepackage{color, colortbl}
\usepackage[binary-units]{siunitx}
\DeclareSIUnit{\bps}{bps}
\usepackage{amsmath,amssymb,amsfonts}
\usepackage{cite}
\usepackage{graphicx}
\def\BibTeX{{\rm B\kern-.05em{\sc i\kern-.025em b}\kern-.08em
    T\kern-.1667em\lower.7ex\hbox{E}\kern-.125emX}}
\begin{document}
\title{Multi-User Reconfigurable Intelligent Surface-Aided Communications Under Discrete Phase Shifts}

\author{\IEEEauthorblockN{Wei Jiang\IEEEauthorrefmark{1}\IEEEauthorrefmark{2} and Hans D. Schotten\IEEEauthorrefmark{2}\IEEEauthorrefmark{1}}
\IEEEauthorblockA{\IEEEauthorrefmark{1}Intelligent Networking Research Group, German Research Center for Artificial Intelligence (DFKI),  Germany\\
  }
\IEEEauthorblockA{\IEEEauthorrefmark{2}Department of Electrical and Computer Engineering, Technische Universit\"at (TU) Kaiserslautern, Germany\\
 }
}


%


\maketitle

\begin{abstract}
This paper focuses on studying  orthogonal and non-orthogonal multiple access in intelligent reflecting surface (IRS)-aided systems. Unlike most prior works assuming continuous phase shifts, we employ the practical setup where only a finite number of discrete phase shifts are available. To maximize the sum rate, active beamforming and discrete reflection need to be jointly optimized. We therefore propose an alternative optimization method to get the optimal continuous phase shifts iteratively, and then quantize each phase shift to its nearest discrete value. The sum-rate maximization of different schemes is theoretically analyzed and numerically evaluated with different numbers of phase-control bits.
\end{abstract}

%
\IEEEpeerreviewmaketitle

\section{Introduction}

Reconfigurable intelligent surface a.k.a intelligent reflecting surface (IRS) has received considerable attention from the research community \cite{Ref_renzo2020smart}. Particularly, IRS is a planar meta-surface composed of massive reflection elements, each of which can independently induce a phase shift to an impinging signal \cite{Ref_wu2019intelligent}. These elements thereby collaboratively realize a smartly reconfigurable radio environment for signal amplification or interference suppression. IRS not only reflects signals in a full-duplex and noise-free way \cite{Ref_renzo2020reconfigurable} but also lowers hardware and energy costs substantially due to the use of passive components. Consequently, it is recognized that IRS can be a key technological enabler for the upcoming sixth-generation (6G) \cite{Ref_jiang2021road, Ref_jiang2021kickoff} system to meet more stringent performance requirements than its predecessor. To exploit the potential, a growing body of literature have investigated different aspects for IRS-aided wireless communications, i.e., reflection optimization design \cite{Ref_wu2019intelligent}, channel estimation \cite{Ref_wang2020channel}, and the joint design of IRS with other wireless technologies, e.g., orthogonal frequency-division multiplexing (OFDM) \cite{Ref_zheng2020intelligent}, multi-input multi-output (MIMO) \cite{Ref_hu2018beyond}, hybrid beamforming \cite{Ref_di2020hybrid}, millimeter-wave \cite{Ref_jiang2022dualbeam}, and Terahertz communications \cite{Ref_ning2021terahertz}.

Prior works on IRS mostly focus on point-to-point communications that considers a base station (BS), a surface, and a single user. Nevertheless, a practical wireless system needs to accommodate many users simultaneously, raising the problem of multiple access.  On the one hand, its disruptive capability in smartly reconfiguring wireless environments enables new paradigms to mitigate multi-user interference. On the other hand, frequency-division approaches suffer from the lack of frequency-selective reflection  due to IRS hardware constraints.  Hence, it is worth clarifying which multiple access technique is more favorable in IRS-aided systems. In \cite{Ref_zheng2020intelligent_COML}, the authors conducted performance comparison of non-orthogonal multiple access (NOMA), frequency-division multiple access (FDMA), and time-division multiple access (TDMA). But this work is preliminary and needs to be deepened because it assumes that: (1) the BS has only a single antenna, (2) the system merely includes two users, and (3) the channel model is too simple.

Therefore, this paper focuses on studying different orthogonal and non-orthogonal multiple access techniques in a more general IRS-aided multi-user MIMO (MU-MIMO) system, where a multi-antenna BS simultaneously serves $K$ users. Unlike most prior works employing continuous phase shifts, we use a more practical setup where only a finite number of discrete phase shifts are available \cite{Ref_wu2020beamforming}. To maximize the sum rate of the multi-user IRS system, active beamforming at the BS and discrete passive reflection at the IRS are required to be jointly optimized. Therefore, we propose a two-step alternative optimization method, which gets the optimal continuous phase shifts first, and then quantizes each phase shift to its nearest discrete value. The sum-rate maximization of different multiple-access schemes is theoretically analyzed and comparatively evaluated through simulations with different numbers of phase-control bits. In addition, a more practical simulation setup consisting of a cell-edge area and a cell-center area is deliberately designed, where the COST-Hata model is applied for large-scale fading, and Rician fading is employed to model the line-of-sight (LOS) path between the BS and IRS.

The remainder of this paper is organized as follows: Section II introduces the model of an IRS-aided MU-MIMO system. Section III analyzes three multiple access schemes. Simulation setup and some examples of numerical results are demonstrated in Section IV. Finally, the conclusions are drawn in Section V.

\section{System Model}
\begin{figure}[!h]
    \centering
    \includegraphics[width=0.42\textwidth]{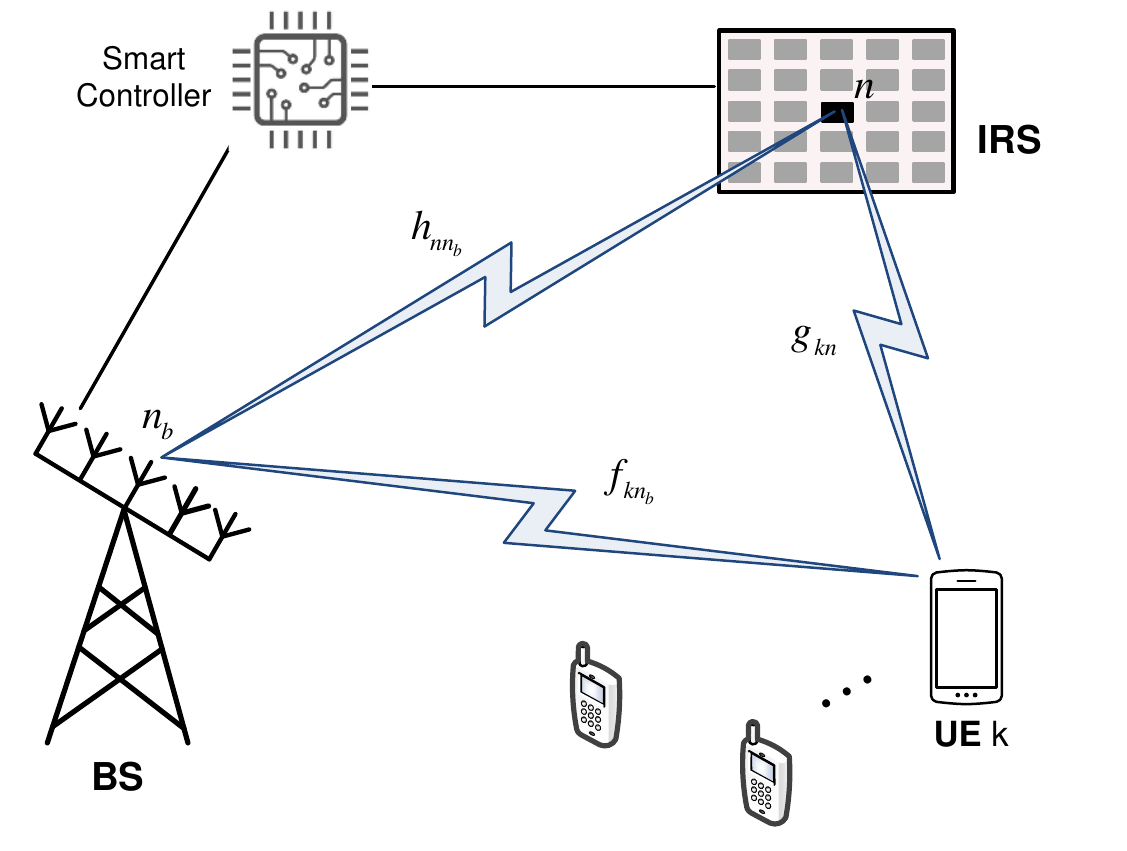}
    \caption{Schematic diagram of an IRS-aided multi-user MIMO system, which is comprised of a multi-antenna BS, $K$ single-antenna UE, and an IRS with $N$ reflecting elements.  }
    \label{diagram:system}
\end{figure}

As illustrated in \figurename \ref{diagram:system}, this paper considers an IRS-assisted MU-MIMO downlink communications system, where an intelligent surface with $N$ reflecting elements is deployed to assist the transmission from an $N_b$-antenna BS to $K$ single-antenna user equipment (UE). Since the direct paths from either the BS or the IRS to UEs may be blocked, the corresponding small-scale fading follows Rayleigh distribution. Consequently, the channel gain between antenna element $n_b\in  \{1,2,\ldots,N_b\}$ and user $k\in \mathcal{K} \triangleq\{1,2,\ldots,K\}$ is a circularly symmetric complex Gaussian random variable with zero mean and variance $\sigma_f^2$, i.e., $f_{kn_b} \sim \mathcal{CN}(0,\sigma_f^2)$. Thus, the channel vectors from the BS and the IRS to the $k^{th}$ UE are denoted by
\begin{align}
    \mathbf{f}_{k}=\Bigl[f_{k1},f_{k2},\ldots,f_{kN_b}\Bigr]^T,
\end{align}
and
\begin{align}
    \mathbf{g}_{k}=\Bigl[g_{k1},g_{k2},\ldots,g_{kN}\Bigr]^T,
\end{align}
respectively, where $g_{kn}\sim \mathcal{CN}(0,\sigma_g^2)$ is the channel gain between IRS element $n\in \mathcal{N} \triangleq \{1,2,\ldots,N\}$ and user $k$.
We write
\begin{equation}
    \mathbf{h}_{n}=[h_{n1},h_{n2},\ldots,h_{nN_b}]^T
\end{equation} to denote the channel vector between the BS and the $n^{th}$ reflecting element. So the channel matrix from the BS to the IRS is expressed as $\mathbf{H}\in \mathbb{C}^{N\times N_b}$, where the $n^{th}$ row of $\mathbf{H}$ equals to $\mathbf{h}_n^T$.  In contrast to randomly distributed and moving UEs, a favourable location is deliberately selected for the IRS to exploit a LOS path of the fixed BS without any blockage, resulting in Rician fading, i.e.,
\begin{equation}\label{EQNIRQ_LSFadingdirect}
    \mathbf{H}=\sqrt{\frac{\Gamma\sigma_h^2}{\Gamma+1}}\mathbf{H}_{LOS} + \sqrt{\frac{\sigma_h^2}{\Gamma+1}}\mathbf{H}_{NLOS},
\end{equation}
with the Rician factor $\Gamma$, the LOS component $\mathbf{H}_{LOS}$,  the multipath component $\mathbf{H}_{NLOS}$ consisting of independent entries that follow $\mathcal{CN}(0,1)$, and the BS-IRS path loss $\sigma_h^2$.

Since the IRS is a passive device, time-division duplexing (TDD) operation is usually adopted to simplify channel estimation. The users send pilot signals in the uplink training and the BS estimates the uplink instantaneous channel state information (CSI), which is used for optimizing downlink data transmission due to channel reciprocity.  To characterize the theoretical performance, the analysis hereinafter is conducted under the assumption that the CSI of all involved channels is perfectly known at the BS, as most prior works \cite{Ref_wu2019intelligent, Ref_renzo2020reconfigurable, Ref_renzo2020smart, Ref_wang2020channel,Ref_zhi2021uplink,Ref_zheng2020intelligent,Ref_hu2018beyond,Ref_di2020hybrid,Ref_jiang2022dualbeam,Ref_ning2021terahertz}. In addition, we assume narrowband communications, where the channels follow frequency-flat block fading. A wideband channel suffering from frequency selectivity can be transformed into a set of narrowband channels through OFDM \cite{Ref_jiang2016ofdm}, making the assumption of flat fading reasonable.

The reflecting surface is equipped with a smart controller, which can adaptively adjust the phase shift of each IRS element in terms of the CSI acquired through periodic channel estimation \cite{Ref_wang2020channel}. Ideally, the coefficient of each reflection element can be continuously adjusted. Denoting the reflection coefficient of the $n^{th}$ element by $c_{n}=\beta_{n} e^{j\theta_{n}}$ yields a continuous phase shift $\theta_{n}\in [0,2\pi)$ and amplitude attenuation $\beta_{n}\in [0,1]$. As revealed by \cite{Ref_wu2019intelligent}, the optimal value of attenuation is $\beta_{n}=1$, $\forall n$ to maximize the received power and simplify hardware implementation, which will therefore not be discussed hereinafter. Although continuous phase shifts are beneficial for optimizing performance, it is practically difficult to implement. Positive-intrinsic-negative (PIN) diodes have been widely adopted for fabricating reflection elements due to fast response time, small reflection loss, and relatively low hardware and energy costs. By adding different biasing voltages, a PIN diode is switched to either ON or OFF state, inducing a phase-shift difference of $\pi$. Finer controlling is realized by integrating multiple diodes. For example, implementing $8$ discrete phase shifts needs at least $\log_28=3$ PIN diodes \cite{Ref_wu2021intelligent}.
Higher resolution imposes not only high hardware cost and design complexity but also the backhaul burden of a wireless or wired link between the BS and the smart controller. Since a surface usually contains a large number of reflecting elements, it is practical to implement only a finite number of discrete phase shifts denoted by $\phi_n$ hereinafter. Let $b$ denote the number of bits for representing $L=2^b$ different levels,
the set of discrete phase shifts can be expressed as
\begin{equation}
    \mathcal{F} = \Bigl\{0, \triangle \phi, 2\triangle \phi, \ldots, (L-1) \triangle \phi \Bigr\}
\end{equation}
with $\triangle \phi=2\pi/L$.

Because of severe path loss, the signals that are reflected by the IRS twice or more are negligible. By ignoring impairments such as channel aging \cite{Ref_jiang2021impactcellfree} and phase noise \cite{Ref_jiang2022impact}, the $k^{th}$ UE observes the received signal
\begin{equation} \label{eqn_systemModel}
    r_k=\sqrt{P_t}\Biggl( \sum_{n=1}^N g_{kn} e^{j\phi_{n}} \mathbf{h}_{n}^T + \mathbf{f}_{k}^T \Biggr) \mathbf{s} + n_k,
\end{equation}
where $\mathbf{s}$ denotes the vector of transmitted signals over the BS antenna array, $P_t$ expresses the transmit power, $n_k$ is additive white Gaussian noise (AWGN) with zero mean and variance $\sigma_n^2$, namely $n_k\sim \mathcal{CN}(0,\sigma_n^2)$. Define
\begin{equation}
    \boldsymbol{\Phi}_0=\mathrm{diag}\Bigl\{e^{j\phi_{1}},e^{j\phi_{2}},\ldots,e^{j\phi_{N}}\Bigr\}, \:\phi_{n}\in \mathcal{F},
\end{equation}
\eqref{eqn_systemModel} can be rewritten in matrix form as
\begin{equation} \label{EQN_IRS_RxSignal_Matrix}
    r_k= \sqrt{P_t}\Bigl(\mathbf{g}_k^T \boldsymbol{\Phi}_0 \mathbf{H} +\mathbf{f}_k^T\Bigr)\mathbf{s} +n_k.
\end{equation}

\section{Orthogonal and Non-Orthogonal Multiple Access}
This section analyzes sum spectral efficiencies of TDMA, FDMA and power-domain NOMA in an IRS-aided MU-MIMO system and presents an alternating method to optimize active beamforming at the BS and discrete passive reflection at the IRS jointly.
\subsection{Time-Division Multiple Access}  This scheme orthogonally divides a radio frame into $K$ time slots. Each user transmits over the entire bandwidth but cyclically accesses its assigned slot. At the header of each frame, the BS processes the uplink training signals to get the CSI, which keeps constant for the whole frame. At the $k^{th}$ slot, the BS applies linear beamforming $\mathbf{w}_k\in \mathbb{C}^{N_b\times 1}$, where $\|\mathbf{w}_k\|^2\leqslant 1$, to send the information-bearing symbol $s_k$ with zero mean and unit variance, i.e., $\mathbb{E}\left[|s_k|^2\right]=1$, intended for a general user $k$.
According to \cite{Ref_zhang2018space}, the switching frequency of PIN diodes can reach 5 megahertz (\si{\mega\hertz}), corresponding to the switching time of \SI{0.2}{\micro\second}, much smaller than the typical channel coherence time on the order of millisecond (\si{\milli\second}). It implies that the set of phase shifts can be adjusted specifically for the active user at each time slot. We write $\phi_{nk}\in \mathcal{F}$ to denote the phase shift of the $n^{th}$ IRS element at slot $k$, and define
\begin{equation}
    \boldsymbol{\Phi}_k=\mathrm{diag}\Bigl\{e^{j\phi_{1k}},e^{j\phi_{2k}},\ldots,e^{j\phi_{Nk}}\Bigr\},\:\:k\in \mathcal{K}.
\end{equation}

\begin{figure}[!t]
    \centering
    \includegraphics[width=0.45\textwidth]{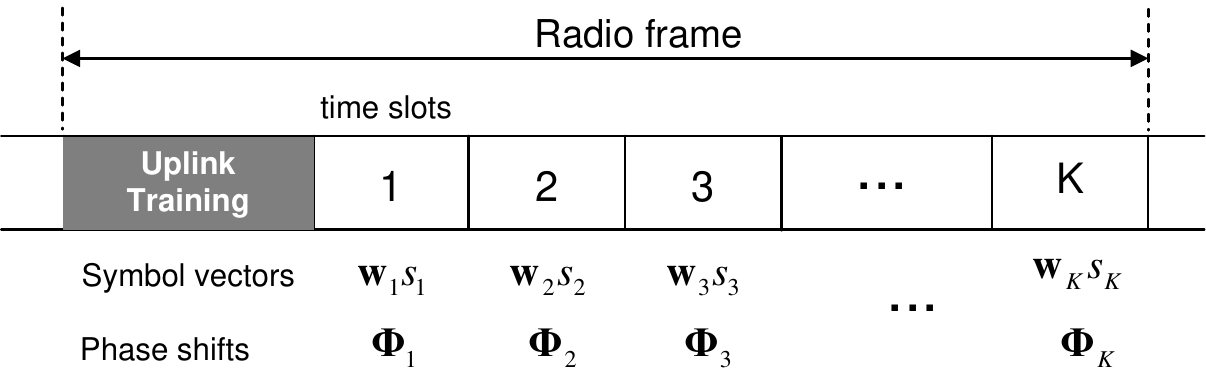}
    \caption{The structure of a TDMA frame.  }
    \label{diagram:TDMA}
\end{figure}
Substituting $\mathbf{s}=\mathbf{w}_k s_k$ and $\boldsymbol{\Phi}_0=\boldsymbol{\Phi}_k$ into \eqref{EQN_IRS_RxSignal_Matrix}, we obtain
\begin{equation}
    r_k= \sqrt{P_d}\Bigl(\mathbf{g}_k^T \boldsymbol{\Phi}_k \mathbf{H} +\mathbf{f}_k^T\Bigr)\mathbf{w}_k s_k +n_k,
\end{equation}
where $P_d$ stands for the power constraint of the BS, and $P_t=P_d$ in TDMA.
The instantaneous signal-to-noise ratio (SNR) of user $k$, i.e.,
\begin{equation} \label{IRS_EQN_spectralEfficiency}
    \gamma_k=\frac{P_d \Bigl|\bigl(\mathbf{g}_k^T \boldsymbol{\Phi}_k \mathbf{H} +\mathbf{f}_k^T\bigr)\mathbf{w}_k\Bigr|^2 }{\sigma_n^2}
\end{equation}
can be maximized by jointly optimizing $\mathbf{w}_k$ and $\boldsymbol{\Phi}_k$, resulting in the following optimization formula
\begin{equation}
\begin{aligned} \label{eqnIRS:optimizationMRTvector}
\max_{\boldsymbol{\Phi}_k,\:\mathbf{w}_k}\quad &  \biggl|\Bigl(\mathbf{g}_k^T \boldsymbol{\Phi}_k \mathbf{H} +\mathbf{f}_k^T\Bigr)\mathbf{w}_k\biggr|^2\\
\textrm{s.t.} \quad & \|\mathbf{w}_k\|^2\leqslant 1\\
  \quad & \phi_{nk}\in \mathcal{F}, \: \forall n\in \mathcal{N},\: \forall k\in \mathcal{K},
\end{aligned}
\end{equation}
which is non-convex because the objective function is not jointly concave with respect to $\boldsymbol{\Phi}_k$ and $\mathbf{w}_k$. To solve this problem, we propose a two-step alternative optimization method, which first gets the optimal continuous phase shifts in an iterative manner, and then quantizes each phase shift to its nearest discrete value.
Correspondingly, the constraint of discrete phase shifts $\phi_{nk}\in \mathcal{F}$ in \eqref{eqnIRS:optimizationMRTvector} needs to be relaxed to continuous phase shifts $\theta_{nk}\in [0,2\pi)$.
Given an initialized transmit vector $\mathbf{w}_k^{(0)}$, \eqref{eqnIRS:optimizationMRTvector} is rewritten as
\begin{equation}  \label{eqnIRS:optimAO}
\begin{aligned} \max_{\boldsymbol{\Theta}_k}\quad &  \biggl|\Bigl(\mathbf{g}_k^T \boldsymbol{\Theta}_k \mathbf{H} +\mathbf{f}_{k}^T\Bigr)\mathbf{w}_k^{(0)}\biggr|^2\\
\textrm{s.t.}  \quad & \theta_{nk}\in [0,2\pi), \: \forall n\in \mathcal{N},\: \forall k\in \mathcal{K},
\end{aligned}
\end{equation}
with $\boldsymbol{\Theta}_k=\mathrm{diag}\left\{e^{j\theta_{1k}},e^{j\theta_{2k}},\ldots,e^{j\theta_{Nk}}\right\}$.
The objective function is still non-convex but it enables a closed-form solution through applying the well-known triangle inequality
\begin{equation}
    \biggl|\Bigl(\mathbf{g}_k^T \boldsymbol{\Theta}_k \mathbf{H} +\mathbf{f}_{k}^T\Bigr)\mathbf{w}_k^{(0)}\biggr| \leqslant \biggl|\mathbf{g}_k^T \boldsymbol{\Theta}_k \mathbf{H} \mathbf{w}_k^{(0)}\biggr| +\biggl|\mathbf{f}_{k}^T\mathbf{w}_k^{(0)}\biggr|.
\end{equation}
The equality achieves if and only if
\begin{equation}
    \arg\left (\mathbf{g}_k^T \boldsymbol{\Theta}_k \mathbf{H} \mathbf{w}_k^{(0)}\right)= \arg\left(\mathbf{f}_{k}^T\mathbf{w}_k^{(0)}\right)\triangleq \varphi_{0k},
\end{equation}
where $\arg(\cdot)$ stands for the phase of a complex scalar.

Define $\mathbf{q}_k=\left[q_{1k},q_{2k},\ldots,q_{Nk}\right]^H$ with $q_{nk}=e^{j\theta_{nk}}$ and  $\boldsymbol{\chi}_k=\mathrm{diag}(\mathbf{g}_k^T)\mathbf{H}\mathbf{w}_k^{(0)}\in \mathbb{C}^{N\times 1}$, we have $\mathbf{g}_k^T \boldsymbol{\Theta}_k \mathbf{H} \mathbf{w}_k^{(0)}=\mathbf{q}_k^H\boldsymbol{\chi}_k\in \mathbb{C} $.
Ignore the constant term $\bigl|\mathbf{f}_{k}^T\mathbf{w}_k^{(0)}\bigr|$, \eqref{eqnIRS:optimAO} is transformed to
\begin{equation}  \label{eqnIRS:optimizationQ}
\begin{aligned} \max_{\boldsymbol{\mathbf{q}_k}}\quad &  \Bigl|\mathbf{q}_k^H\boldsymbol{\chi}_k\Bigl|\\
\textrm{s.t.}  \quad & |q_{nk}|=1, \: \forall n\in \mathcal{N},\: \forall k\in \mathcal{K},\\
  \quad & \arg(\mathbf{q}_k^H\boldsymbol{\chi}_k)=\varphi_{0k}.
\end{aligned}
\end{equation}
An IRS should be tuned such that the reflected signals and the signals over the direct link are phase-aligned to achieve coherent combining. As a result, the solution for \eqref{eqnIRS:optimizationQ} is
\begin{equation} \label{eqnIRScomplexityQ}
    \mathbf{q}^{(1)}_k=e^{j\left(\varphi_{0k}-\arg(\boldsymbol{\chi}_k)\right)}=e^{j\left(\varphi_{0k}-\arg\left( \mathrm{diag}(\mathbf{g}_k^T)\mathbf{H}\mathbf{w}_k^{(0)}\right)\right)}.
\end{equation}
Accordingly,
\begin{equation} \nonumber \label{IRSeqnOptimalShift}
    \theta_{nk}^{(1)}=\varphi_{0k}-\arg\left(g_{nk}\right)-\arg\left(\mathbf{h}_{n}^T\mathbf{w}_k^{(0)}\right).
\end{equation}
Once the set of phase shifts at the first iteration, i.e., $    \boldsymbol{\Theta}^{(1)}_k=\mathrm{diag}\left\{e^{j\theta_{1k}^{(1)}},e^{j\theta_{2k}^{(1)}},\ldots,e^{j\theta_{Nk}^{(1)}}\right\}$ are determined, the optimization is alternated to update $\mathbf{w}_k$. The BS can apply maximal-ratio transmission (MRT) to maximize the strength of a desired signal, resulting in
\begin{equation}  \label{EQN_IRS_TXBF}
    \mathbf{w}_k^{(1)} = \frac{\left(\mathbf{g}_k^T \boldsymbol{\Theta}^{(1)}_k \mathbf{H} +\mathbf{f}_{k}^T\right)^H}{\left\|\mathbf{g}_k^T \boldsymbol{\Theta}_k^{(1)} \mathbf{H} +\mathbf{f}_{k}^T\right\|}.
\end{equation}

After the completion of the first iteration, the BS gets $\boldsymbol{\Theta}^{(1)}_k$ and $\mathbf{w}_k^{(1)}$, which serve as the initial input for the second iteration to derive $\boldsymbol{\Theta}^{(2)}_k$ and $\mathbf{w}_k^{(2)}$.
This process iterates until the convergence is achieved with the optimal transmit vector $\mathbf{w}_k^{\star}$ and continuous phase shifts $\boldsymbol{\Theta}_k^{\star}=\mathrm{diag}\left\{e^{j\theta_{1k}^{\star}},e^{j\theta_{2k}^{\star}},\ldots,e^{j\theta_{Nk}^{\star}}\right\}$. Lastly, each phase shift is quantized into a finite number of phase-control bits before sending to the smart controller through a wireless or wired link, applying a mid-tread uniform quantizer
\begin{equation}
    \phi_{nk}^\star = \triangle \phi \left \lfloor  \frac{\theta_{nk}^\star}{\triangle \phi } + 0.5\right\rfloor,
\end{equation}
where the notation $\lfloor \cdot \rfloor$  denotes the floor function.

Substituting $\mathbf{w}_k^{\star}$ and  $\boldsymbol{\Phi}_k^{\star}=\mathrm{diag}\left\{e^{j\phi_{1k}^{\star}},e^{j\phi_{2k}^{\star}},\ldots,e^{j\phi_{Nk}^{\star}}\right\}$ into \eqref{IRS_EQN_spectralEfficiency}, we can derive the maximal spectral efficiency of user $k$ as
\begin{equation} \label{IRS_EQN_TDMA_SE}
    C_k=\frac{1}{K}\log\left(1+\frac{P_d \Bigl|\bigl(\mathbf{g}_k^T \boldsymbol{\Phi}_k^\star \mathbf{H} +\mathbf{f}_k^T\bigr)\mathbf{w}_k^\star \Bigr|^2 }{\sigma_n^2} \right).
\end{equation}
Thereby, the sum rate maximization of the TDMA-based IRS system can be computed by $C_{tdma}=\sum_{k=1}^K C_k$.

\subsection{Frequency-Division Multiple Access} In FDMA, the system bandwidth is divided along the frequency axis into $K$ orthogonal subchannels. Each user occupies a dedicated subchannel over the entire time. The BS employs linear beamforming $\mathbf{w}_k$ to transmit $s_k$ over the $k^{th}$ subchannel with equally-allocated power $P_t=P_d/K$.
In contrast to TDMA, where the IRS phase shifts can be dynamically adjusted in different slots,  the surface can be optimized only for a particular user, whereas other users suffer from phase-unaligned reflection. That is because the hardware limitation of IRS passive elements, which can be fabricated in \textit{time-selective} rather than \textit{frequency-selective}.
Without losing generality, we suppose the FDMA system optimizes the IRS to aid  the signal transmission of user $\hat{k}$, the optimal parameters $\boldsymbol{\Phi}^{\star}_{\hat{k}}$ and $\mathbf{w}_{\hat{k}}^{\star}$ can be derived using the same alternating optimization as that of TDMA.
Thus, the achievable spectral efficiency of user $\hat{k}$ is
\begin{equation}
    C_{\hat{k}}=\frac{1}{K}\log\left(1+\frac{P_d/K \left|\bigl(\mathbf{g}_{\hat{k}}^T \boldsymbol{\Phi}_{\hat{k}}^\star \mathbf{H} +\mathbf{f}_{\hat{k}}^T\bigr)\mathbf{w}_{\hat{k}}^\star \right|^2 }{\sigma_n^2/K} \right),
\end{equation}
where the factor $1/K$ is due to the fact that each user occupies only $1/K$ of the total bandwidth.
Once the phase shifts of the surface are completely adjusted for $\hat{k}$,  what the remaining $K-1$ users, denoted by $i\in \mathcal{K}-\left\{\hat{k}\right\}$, can do is to only optimize their respective active beamforming based on the fixed reflection $\boldsymbol{\Phi}_{\hat{k}}$. For user $i$, the MRT is given by
\begin{equation}  \label{EQN_IRS_matchedFilter}
    \mathbf{w}_{i}^{\star} = \frac{\Bigl(\mathbf{g}_{i}^T \boldsymbol{\Phi}^{\star}_{\hat{k}} \mathbf{H} +\mathbf{f}_{i}^T\Bigr)^H}{\Bigl\|\mathbf{g}_{i}^T \boldsymbol{\Phi}_{\hat{k}}^{\star} \mathbf{H} +\mathbf{f}_{i}^T\Bigr\|}.
\end{equation}
Then, the sum-rate maximization of the FDMA IRS system is
 \begin{align} \nonumber
     C_{fdma}&= \frac{1}{K}\log\left(1+\frac{P_d \Bigl|\bigl(\mathbf{g}_{\hat{k}}^T \boldsymbol{\Phi}_{\hat{k}}^\star \mathbf{H} +\mathbf{f}_{\hat{k}}^T\bigr)\mathbf{w}_{\hat{k}}^\star\Bigr|^2 }{\sigma_n^2} \right)  \\ \nonumber
     &+\sum_{i\in \mathcal{K}-\{\hat{k}\}} \frac{1}{K}\log\left(1+\frac{P_d \Bigl|\bigl(\mathbf{g}_{i}^T \boldsymbol{\Phi}_{\hat{k}}^\star \mathbf{H} +\mathbf{f}_{i}^T\bigr)\mathbf{w}_{i}^\star\Bigr|^2 }{\sigma_n^2} \right)\\
     &= \frac{1}{K}\log\left(1+\frac{P_d \Bigl|\bigl(\mathbf{g}_{\hat{k}}^T \boldsymbol{\Phi}_{\hat{k}}^\star \mathbf{H} +\mathbf{f}_{\hat{k}}^T\bigr)\mathbf{w}_{\hat{k}}^\star\Bigr|^2 }{\sigma_n^2} \right)  \\ \nonumber
     &+\sum_{i\in \mathcal{K}-\{\hat{k}\}} \frac{1}{K}\log\left(1+ \bigl\|\mathbf{g}_{i}^T \boldsymbol{\Phi}_{\hat{k}}^\star \mathbf{H} +\mathbf{f}_{i}^T \bigr\|^2 \frac{P_d }{\sigma_n^2} \right).
  \end{align}

\subsection{Non-Orthogonal Multiple Access}
OMA can mitigate multi-user interference among orthogonally multiplexed users to facilitate low-complexity receiver, but it cannot achieve the sum-rate capacity of a multi-user system. Using superposition coding and successive interference cancellation (SIC), NOMA serves more than one user over each orthogonal resource unit. The BS superimposes $K$ symbols into a composite signal
\begin{equation} \label{EQNIRQ_compositeSignal}
    \mathbf{s}=\sum_{k=1}^K \sqrt{\alpha_k } \mathbf{w}_{k} s_k,
\end{equation}
where  $\alpha_k$ represents the power allocation coefficient subjecting to $\sum_{k=1}^K\alpha_k\leqslant 1$. NOMA is a kind of power-domain multiple access, where more power is allocated to the users with smaller channel gain.

As FDMA, we suppose the system optimizes the IRS to aid the signal transmission of user $\hat{k}$.
The optimal parameters $\boldsymbol{\Phi}^{\star}_{\hat{k}}$ and $\mathbf{w}_{\hat{k}}^{\star}$ can be derived using the alternating optimization as in TDMA. Once the phase shifts of the surface are set to $\boldsymbol{\Phi}^{\star}_{\hat{k}}$,  a non-IRS-aided user $i\in \mathcal{K}-\left\{\hat{k}\right\}$ can obtain its optimal active beamforming $\mathbf{w}_{k}^{\star}$ as \eqref{EQN_IRS_matchedFilter}.
  Substitute \eqref{EQNIRQ_compositeSignal} into \eqref{EQN_IRS_RxSignal_Matrix} to yield the observation of a typical user $k$ (including $\hat{k}$ and $i\neq \hat{k}$) as
\begin{align} \nonumber
    r_k&= \sqrt{P_d}\Bigl(\mathbf{g}_k^T \boldsymbol{\Phi}_{\hat{k}}^\star \mathbf{H} +\mathbf{f}_k^T\Bigr) \sum_{k'=1}^K \sqrt{\alpha_{k'}} \mathbf{w}_{k'}s_{k'} +n_k\\
    &= \underbrace{\sqrt{\alpha_{k}P_d}\Bigl(\mathbf{g}_k^T \boldsymbol{\Phi}_{\hat{k}}^\star \mathbf{H} +\mathbf{f}_k^T\Bigr) \mathbf{w}_{k}s_{k}}_{\text{Desired\:signal}}\\ \nonumber
    &+ \underbrace{\sqrt{P_d}\Bigl(\mathbf{g}_k^T \boldsymbol{\Phi}_{\hat{k}}^\star \mathbf{H} +\mathbf{f}_k^T\Bigr) \sum_{k'=1,k'\neq k}^K \sqrt{\alpha_{k'}} \mathbf{w}_{k'}s_{k'}}_{\text{Multi-user\:interference}}+n_k.
\end{align}

The optimal order of interference cancellation is detecting the user with the weakest channel gain to the user with the strongest channel gain. We write $\rho_k=(\mathbf{g}_k^T \boldsymbol{\Phi}^{\star}_{\hat{k}} \mathbf{H} +\mathbf{f}_k^T)\mathbf{w}_k^\star$, $\forall k$ to denote the effective gain of the combined channel for user $k$.
Without loss of generality, assume that user $1$ has the largest combined channel gain, and user $K$ is the weakest, i.e.,
\begin{equation}
   \| \rho_1 \|^2 \geqslant \|\rho_2\|^2 \geqslant \ldots \geqslant \|\rho_K\|^2.
\end{equation}
 With this order, each NOMA user decodes $s_K$ first, and then subtracts its resultant component from the received signal. As a result, a typical user $k$ after the first SIC iteration gets
\begin{equation}
    \tilde{r}_k = r_k - \rho_k \sqrt{\alpha_K P_d}  s_K = \rho_k\sum_{k=1}^{K-1} \sqrt{\alpha_k P_d} s_k+n_k,
\end{equation}
assuming error-free detection and perfect channel knowledge.
In the second iteration, the user decodes $s_{K-1}$ using the remaining signal $ \tilde{r}_k$. The cancellation iterates until each user gets the symbol intended for it. Consequently, the received SNR for user $k$ is
\begin{equation}
    \gamma_k=\frac{\|\rho_k\|^2 \alpha_k P_d}{ \|\rho_k\|^2 \sum_{k'=1}^{k-1}\alpha_{k'} P_d + \sigma_n^2 },
\end{equation}
resulting in the achievable rate of $
    R_k= \log \left ( 1+\gamma_k \right)$.
The maximized sum rate of IRS-aided NOMA transmission is computed by
\begin{equation}
    C_{noma}=\sum_{k=1}^K \log \left ( 1+\frac{\|\rho_k\|^2\alpha_k P_d}{\|\rho_k\|^2 \sum_{k'=1}^{k-1}\alpha_{k'} P_d + \sigma_n^2 }   \right).
\end{equation}


\section{Simulation Results}

\begin{figure}[!ht]
    \centering
    \includegraphics[width=0.42\textwidth]{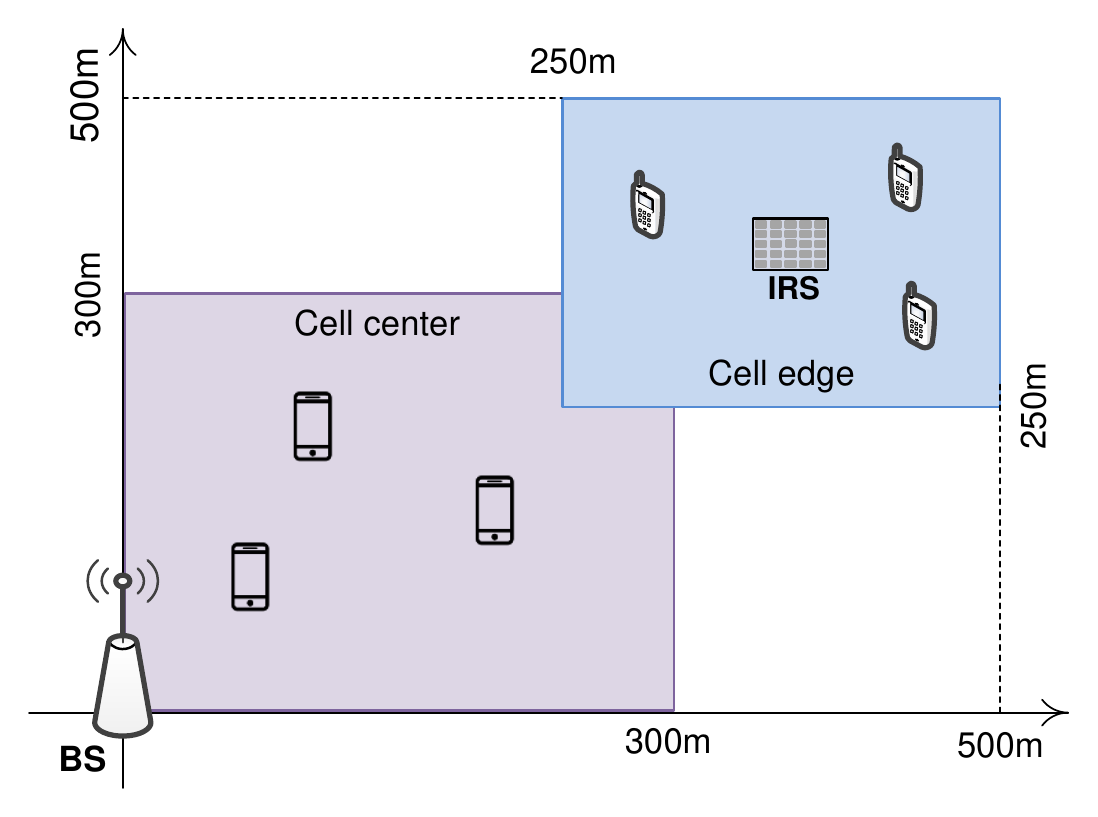}
    \caption{Simulation scenario of a multi-user IRS system, where the cell coverage is comprised of a cell-center area and a cell-edge area. }
    \label{diagram:Simulation}
\end{figure}

This section provides our simulation scenario and illustrates some representative numerical results to compare the performance of different multiple-access techniques in terms of achievable spectral efficiency.
As shown in \figurename \ref{diagram:Simulation}, we consider the cell coverage consisting of a cell-center area and a cell-edge area. The BS equipped with $N_b=16$ antennas is located at the original point $(0,0)$ of the coordinate system, while a surface containing $N=200$ reflecting elements is installed at the center of the cell-edge area, with the coordinate $(375\si{\meter},375\si{\meter})$.  Half of the users are far users that distribute randomly over the cell-edge area, while the other half of users are near users that distribute randomly over the cell-center area. The maximum transmit power of BS is $P_d=20\mathrm{W}$ over a signal bandwidth of $B_w=20\mathrm{MHz}$, conforming with the 3GPP LTE specification.  The variance of white noise  is figured out by $\sigma_n^2=\kappa\cdot B_w\cdot T_0\cdot N_f$ with the Boltzmann constant $\kappa$, temperature $T_0=290 \mathrm{Kelvin}$, and the noise figure  $N_f=9\mathrm{dB}$.
The variances $\sigma_f^2$ and $\sigma_g^2$ mean distance-dependent large-scale fading, which is computed by
$10^\frac{\mathcal{P}+\mathcal{S}}{10}$ with path loss $\mathcal{P}$ and shadowing fading $\mathcal{S}\sim \mathcal{N}(0,\sigma_{sd}^2)$, which generally has a standard derivation $\sigma_{sd}=8\mathrm{dB}$. We use the COST-Hata model, as depicted in \cite{Ref_jiang2021cellfree}, to calculate the path loss with the break points $d_0=10\mathrm{m}$ and $d_1=50\mathrm{m}$, carrier frequency $1.9\mathrm{GHz}$, the height of BS or IRS $15\mathrm{m}$, and the height of UE $1.65\mathrm{m}$. In contrast, the path loss for the LOS channel between the BS and IRS can be computed by $L_0/d^{-\alpha}$,
where $L_0=\SI{-30}{\decibel}$ is the path loss at the reference distance of \SI{1}{\meter}, $d$ represents the propagation distance, $\alpha=2$ means the path loss exponent, and the Rician factor $K=5$.

\begin{figure*}[!t]
\centerline{
\subfloat[]{
\includegraphics[width=0.45\textwidth]{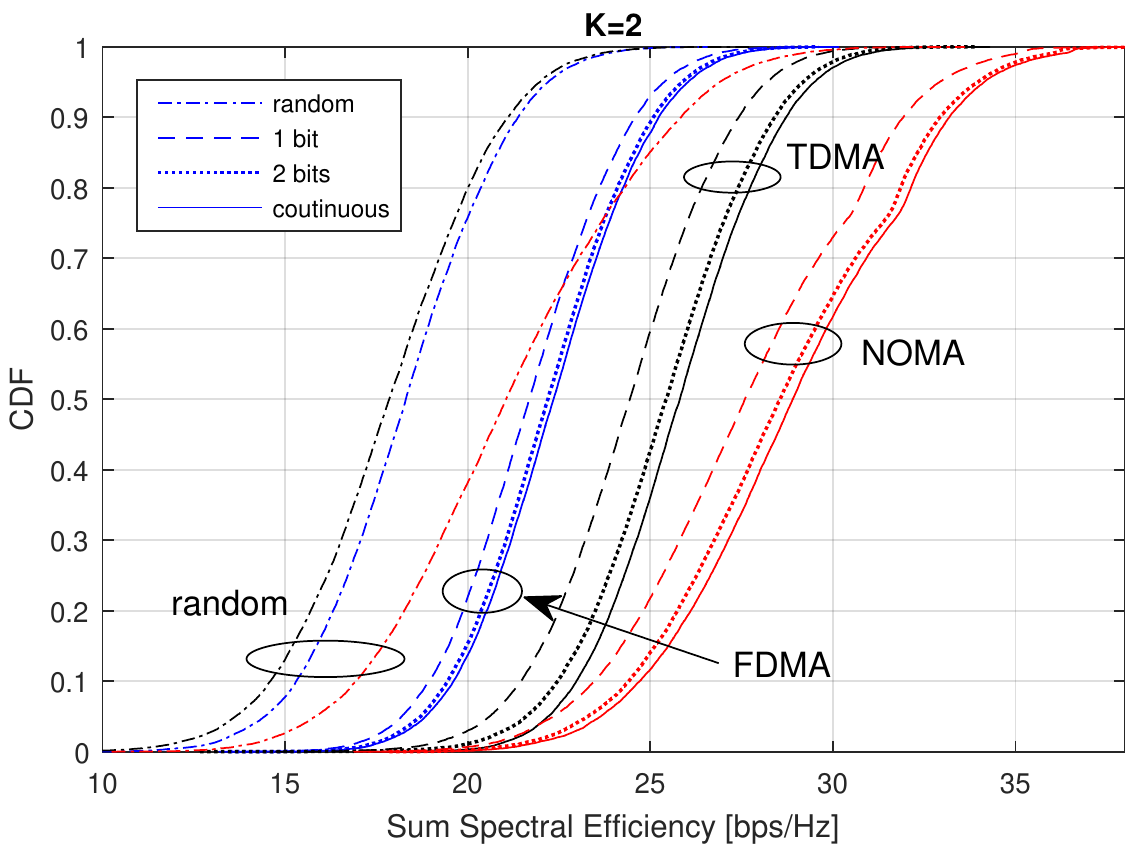}
\label{Fig_results1}
}
\hspace{5mm}
\subfloat[]{
\includegraphics[width=0.375\textwidth]{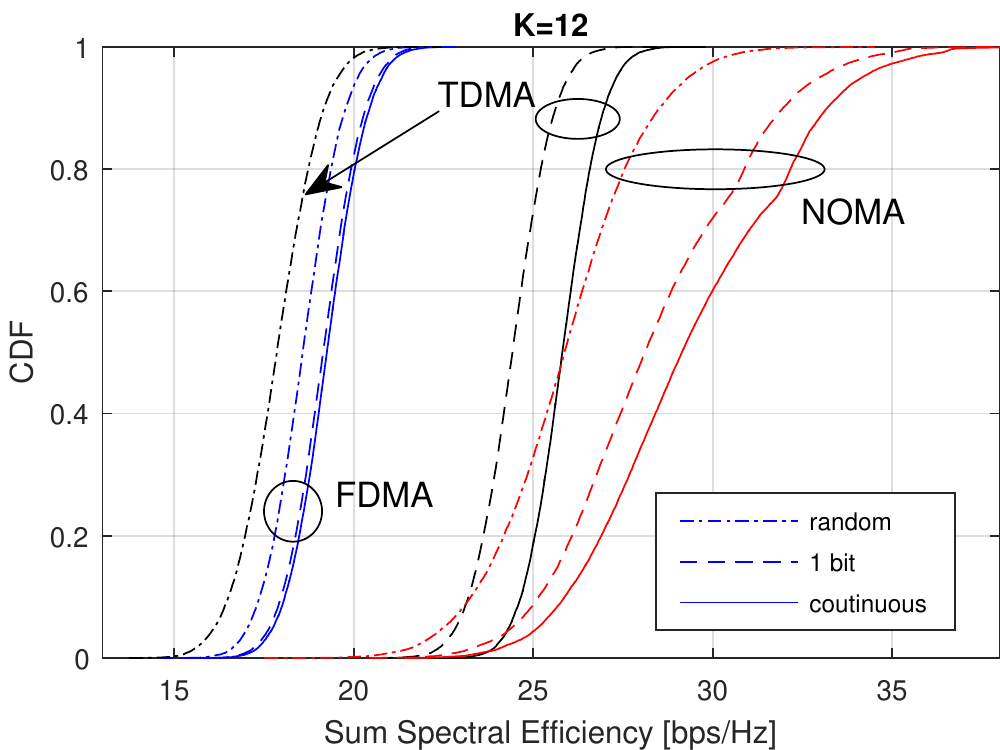}
\label{Fig_results2}
}}
\caption{Performance comparison of FDMA, TDMA, and NOMA in an IRS-aided MU-MIMO system using continuous, discrete, and random phase shifts: (a) CDFs in terms of the sum rate with two users, and (b) CDFs in terms of the sum rate with twelve users. }
\label{Fig_Result}
\end{figure*}

Cumulative distribution functions (CDFs) of the sum spectral efficiency in bits-per-second per hertz (\si{\bps\per\hertz^{}}) achieved by the IRS-aided multi-user MIMO system is used as the performance metric. Numerical results of FDMA, TDMA, and NOMA when phase shifts are \textit{continuous},  \textit{discrete} with $b=1$ and $b=2$ phase-control bits, and \textit{random} by setting $\theta_n\in[0,2\pi)$, $\forall n$ randomly are provided for comparison. In our simulations, the number of iterations for alternating optimization is set to three, which is enough for the convergence of optimization.
\figurename \ref{Fig_results1} shows the CDF curves in the case of $K=2$ users, consisting of a cell-center user and a cell-edge user.
Using random phase shifts, the TDMA or FDMA scheme can achieve the $95\%$-likely spectral efficiency, which is usually applied to measure the performance of cell-edge users, of around \SI{14}{\bps\per\hertz^{}}. As we expected, NOMA is superior to OMA due to the use of advanced signal processing (i.e., superposition coding and SIC). To be specific, NOMA boosts the $95\%$-likely spectral efficiency to \SI{15.83}{\bps\per\hertz^{}}, amounting to a performance growth of over $10\%$.

It is observed that smartly phase control brings significant performance improvement. The $95\%$-likely spectral efficiencies of FDMA using discrete phase shifts of 1 bit and 2 bits, and continuous phase shifts grow to \SI{18.19}{\bps\per\hertz^{}}, \SI{18.71}{\bps\per\hertz^{}}, and \SI{18.85}{\bps\per\hertz^{}}, respectively. Compared with TDMA, which achieves  $95\%$-likely spectral efficiency of approximately \SI{20.6}{\bps\per\hertz^{}}, \SI{21.6}{\bps\per\hertz^{}}, and \SI{22.19}{\bps\per\hertz^{}} using 1 bit, 2 bits, and continuous phase shifts, there is a loss of approximately \SI{3}{\bps\per\hertz^{}}. That is because time-selective IRS elements can aid both users optimally in TDMA-IRS by dynamically changing the phase shifts in different slots, whereas the cell-center user in FDMA-IRS suffers from phase-unaligned reflected signals due to the lack of frequency-selective IRS reflection. As in the case of random phase shifts, NOMA still outperforms the two OMA schemes, achieving the best performance of \SI{22.35}{\bps\per\hertz^{}}, \SI{23.29}{\bps\per\hertz^{}}, and \SI{23.64}{\bps\per\hertz^{}}, respectively, in 1 bit, 2 bits, and continuous phase shifts.
In addition, we also illustrate the numerical results of these schemes in the case of $K=12$ users. As we can see in \figurename \ref{Fig_results2},  similar conclusions can be drawn from their performance comparison. For three multiple access schemes, only $b=2$ phase-control bits are enough for achieving near-optimal performance as that of continuous phase shifts. It is a very encouraging result for both the theoretical research and practical implementation of IRS-aided wireless communications.

\section{Conclusion}
This paper studied the performance of different multiple access techniques, including TDMA, FDMA, and NOMA, in IRS-aided multi-user MIMO systems under a practical setup where only a finite number of discrete phase shifts are available at each reflecting element.
To jointly optimize active beamforming and discrete reflection, we propose a two-step alternative optimization method, which gets the optimal continuous phase shifts first, and then quantizes each phase shift to its nearest discrete value. The sum-rate maximization of different multiple-access schemes is theoretically analyzed and comparatively evaluated through simulations with different numbers of phase-control bits. The results revealed that using only $2$ phase-control bits are sufficient for achieving near-optimal performance as continuous phase shifts. It is an inspiring result for the theoretical research and practical implementation of IRS-aided wireless communications.





%

\bibliographystyle{IEEEtran}
\bibliography{IEEEabrv,Ref_CQR}

\end{document}